\documentclass[12pt]{iopart}

\usepackage{graphicx}
\usepackage{siunitx}
\usepackage{multirow}

\usepackage{array}
\newcolumntype{L}{>{$}l<{$}} 

\providecommand{\keywords}[1]
{
  \small	
  Keywords: #1
}

\begin{document}

\title{Accurate laboratory testing of low-frequency triaxial vibration sensors under various environmental conditions}
\author[T. Shimoda {\it et al.}]{Tomofumi Shimoda,  Wataru Kokuyama, Hideaki Nozato}

\address{National Metrology Institute of Japan, National Institute of Advanced Industrial Science and Technology, 1-1-1 Umezono, Tsukuba, Ibaraki 305-8563, Japan}
\ead{tomofumi.shimoda@aist.go.jp}
\vspace{10pt}

\begin{abstract}
Triaxial vibration sensor are widely used used in various application.
Recently, low-cost sensors based on micro electro mechanical system (MEMS) technology are also becoming more widely adopted.
However, their measurement accuracy can be affected by environmental factors such as temperature.
In this study, we developed an environmental testing system integrated with a triaxial vibration exciter.
The system can reproduce long-stroke, low-frequency triaxial vibrations---such as those caused by huge earthquakes---under temperatures ranging from $-30~^\circ\mathrm{C}$ to $+80~^\circ\mathrm{C}$.
Using this system, the measurement accuracy of vibration sensors can be evaluated under different environmental conditions.
The system provides highly accurate reference measurements using a laser interferometer and reference accelerometers that are primarily calibrated within the system.
The overall accuracy of the reference vibration measurement is estimated to be approximately 0.23~\%.
Based on these reference measurements, we investigated the accuracy of earthquake observations using a MEMS accelerometer as a demonstration.
The system configuration and testing procedures are presented in this paper.
\end{abstract}


\section{Introduction} \label{sec:introduction}
Vibration sensors, such as accelerometers and seismometers, are devices that convert input vibrations into electrical signals using mechanical components and electronic circuits. 
Since the mechanical and electrical properties of these sensors are generally influenced by environmental conditions, their measurement characteristics---such as sensitivity---can also be affected.
This change in performance may lead to reduced measurement accuracy under certain environmental conditions. 
In particular, special attention is required when vibration measurements are conducted in harsh environments, such as field applications.
Therefore, vibration sensors should be tested under various conditions to ensure reliable measurement accuracy.

A method for testing the temperature dependence of accelerometers has been studied \cite{Martin2017,Zhu2018} and standardized in ISO 16063-34~\cite{ISO16063-34}. 
In this method, the end of a vibration exciter to which the sensor under test (SUT) is mounted is inserted into a thermostatic chamber.
The vibration is then applied to the SUT, and its output signal is compared with a reference laser interferometer. 
However, this method is limited to single-axis testing.
Direct evaluation of triaxial vibration measurement accuracy under varying environmental conditions has not been feasible, as it requires applying triaxial vibrations to the SUT. 
Although commercially available systems using triaxial vibration exciters exist, they are primarily designed for vibration durability tests and therefore lack accurate references for performance evaluation. 
Furthermore, due to the limited stroke of these exciters, such systems cannot reproduce long-stroke vibrations, such as those caused by large earthquakes.

In this study, we developed a triaxial environmental testing system to evaluate the accuracy of triaxial vibration measurements. 
The proposed system integrates a thermostatic chamber with a triaxial vibration exciter. 
Unlike the ISO 16063-34~\cite{ISO16063-34} method, where the exciter table is inserted into the chamber, our approach places the chamber on the exciter table, enabling compatibility with long-stroke vibrations. 
As a result, the system can simulate large earthquakes while maintaining controlled environmental conditions for the SUT. 
The SUT output is compared with reference servo accelerometers, which provide vibration measurements traceable to national standards. 
In this paper, we describe the testing methodology and present a demonstration using a MEMS accelerometer.

The remainder of this paper is organized as follows:
Section~\ref{sec:system} introduces the configuration of the measurement system;
Section~\ref{sec:primary} describes the calibration of the reference servo accelerometers at various temperatures;
Section~\ref{sec:triaxial} demonstrates an earthquake observation test as an example of triaxial vibration measurement;
and Section~\ref{sec:conclusion} discusses the results and concludes the study.

\section{Measurement system}	\label{sec:system}
\begin{figure}[tb]
	\begin{center}
	\begin{minipage}{0.8\columnwidth}
		\centering
		\includegraphics[width=\columnwidth]{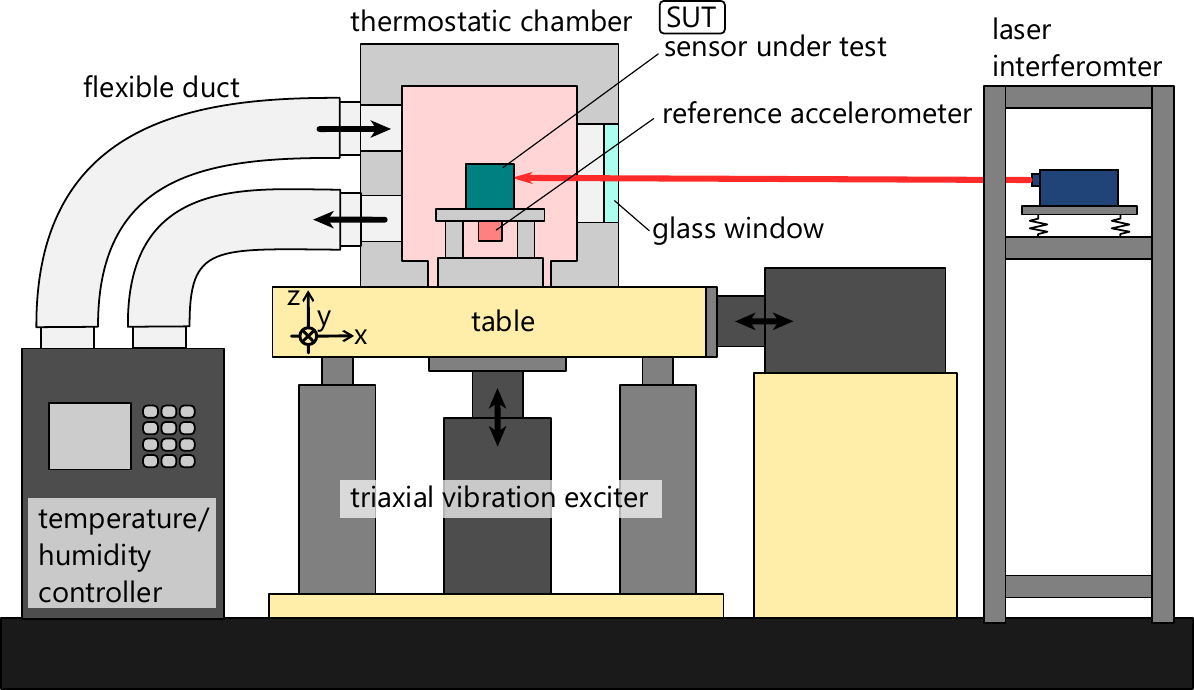}
	\end{minipage}\\
	\vspace{5mm}
	\begin{minipage}{0.5\columnwidth}
		\centering
		\includegraphics[width=\columnwidth]{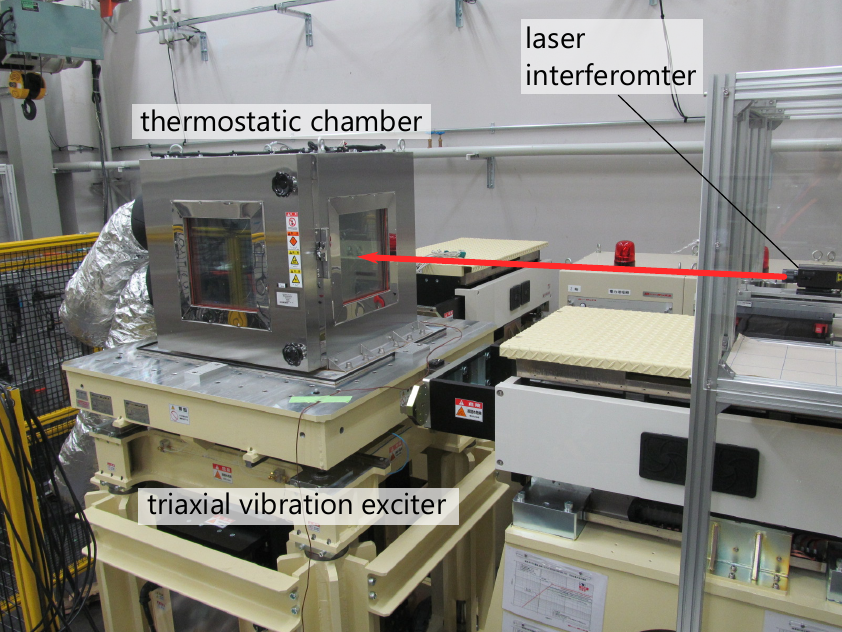}
    	\end{minipage}
	\caption{Schematic diagram (upper) and photo (lower) of the triaxial environmental test system (TETS).}
	\label{fig:system}
	\end{center}
\end{figure}

\begin{table}[b]
\begin{center}
\caption{Specification of the triaxial environmental test system (TETS).}\label{table:spec}
\begin{tabular}{ll}
	\hline
	Vibration exciter	& \\ 
	\hspace{10pt} frequency	range						& 0.1~Hz to 100~Hz \\
	\hspace{10pt} stroke (horizontal, p-p) 			& 400~mm \\
	\hspace{10pt} stroke (vertical, p-p) 				& 100~mm \\
	\hspace{10pt} max. acceleration (horizontal) 	& 24.5~m/s$^2$ \\
	\hspace{10pt} max. acceleration (vertical) 		& 19.2~m/s$^2$ \\ \hline
	Thermostatic chamber	& \\
	\hspace{10pt} inner dimension						& 500~mm$\times$500~mm$\times$500~mm  \\
	\hspace{10pt} temperature range 					& $-30~^\circ$C to $+80~^\circ$C \\
	\hspace{10pt} humidity range 						& 30~\% to 95~\% \\ \hline
\end{tabular}
\end{center}
\end{table}

Figure~\ref{fig:system} shows a schematic diagram of the developed triaxial environmental test system (TETS). 
A thermostatic chamber is mounted on the table of a triaxial vibration exciter. 
The specifications of each component are summarized in Table~\ref{table:spec}. 
The chamber is large enough to accommodate most vibration sensors, including large seismometers. 
The sensor under test is placed inside the chamber, where temperature and humidity are controlled by airflow from the controller through flexible ducts. 
Vibrations are then applied to the entire chamber, including the SUT. 
Unlike the ISO 16063-34 method \cite{ISO16063-34}, where the exciter armature is inserted into a fixed thermostatic chamber, our configuration vibrates the chamber itself along with the SUT. 
This design enables the application of long-stroke triaxial vibrations, which is one of the key features of the system.

The system incorporates two reference measurement devices: a laser interferometer and reference accelerometers. 
Although the laser interferometer provides an absolute reference without calibration, it is unsuitable for measuring long-stroke triaxial vibrations because it requires a large reflective surface, which can degrade accuracy due to structural resonance or imperfect surface flatness. 
Therefore, we adopted a two-step approach for triaxial vibration testing.

First, the reference accelerometers are calibrated using the laser interferometer in a process known as primary calibration, following ISO 16063-11 \cite{ISO16063-11}. 
Each accelerometer is mounted along the x-axis, and a sinusoidal vibration is applied. 
The amplitude of the accelerometer output is compared with that of the interferometer signal to calculate complex sensitivity. 
By repeating this calibration at different temperatures, the temperature dependence of sensitivity is determined. 
Since the inteference of the laser occurs outside the chamber, its measurements are unaffected by the chamber temperature, ensuring that the calibration reflects only the accelerometer’s temperature dependence. 
This primary calibration is performed for three reference accelerometers.

Next, the calibrated reference accelerometers are placed inside the chamber to measure triaxial reference vibrations. 
Because their temperature-dependent sensitivities have been determined in advance, accurate triaxial reference measurements can be obtained at arbitrary temperatures. 
By comparing the SUT output waveform with the reference, measurement accuracy is evaluated.
For example, characteristic parameters such as maximum acceleration, maximum displacement, and seismic intensity scale, can be quantitatively assessed.
Notably, seismic intensity scale calculations require triaxial excitation.

\section{Primary calibration of reference accelerometers}	\label{sec:primary}
\begin{figure}
	\begin{center}
	\includegraphics[width=0.5\columnwidth]{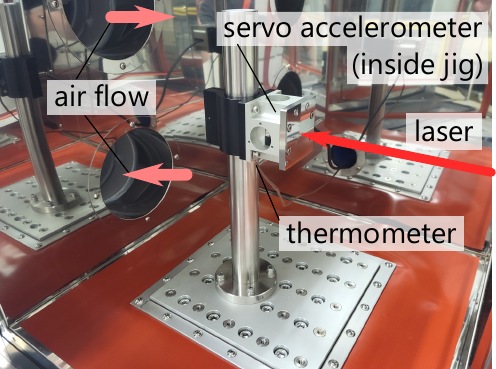}
	\caption{A measurement setup of the servo accelerometer for primary calibration.}
	\label{fig:setup_servo}
	\end{center}
\end{figure}
In this study, three servo accelerometers were calibrated at temperatures ranging from $-20~^\circ$C to $+75~^\circ$C (Section~\ref{sec:primary}). 
Each accelerometer is referred to as “Servo 1,” “Servo 2,” and “Servo 3.” 
The SUT was placed inside the chamber, as shown in Figure~\ref{fig:setup_servo}.

\begin{figure}
	\begin{center}
	\includegraphics[width=0.5\columnwidth]{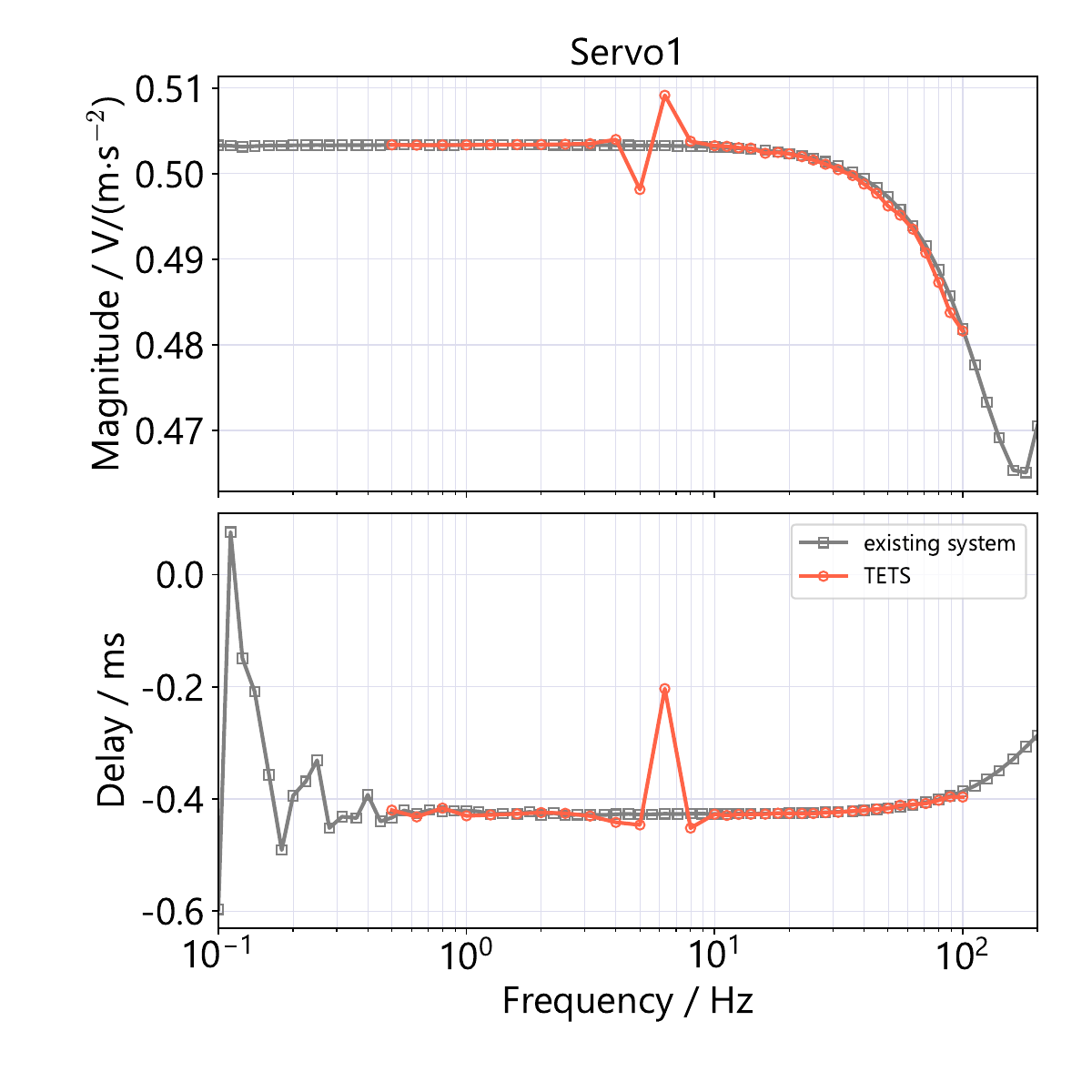}
	\caption{Comparison of primary calibration results using existing system \cite{Shimoda2023} and TETS for Servo 1 at room temperature (23~$^\circ$C).}
	\label{fig:comparison_VChet}
	\end{center}
\end{figure}
First, calibration was performed at room temperature (23~$^\circ$C), and the results were compared with those obtained using an existing calibration system \cite{Shimoda2023}. 
Since the validity of the existing system has been confirmed in previous studies, any differences between the two calibration results were attributed to characteristics of the TETS, such as mechanical resonance of its components. 
The comparison results are shown in Figure~\ref{fig:comparison_VChet}. 
For the complex sensitivity $S$, its magnitude $|S|$ and delay $\arg(S)/(2\pi f)$ are plotted. 
A deviation was observed around 5~Hz, which was identified as the resonance frequency of the frame for the laser interferometer. 
Since the frame was located outside the chamber, the observed deviation is expected to be independent from the temperature at the SUT.
Therefore, subsequent calibration results were corrected to match the sensitivity at 23~$^\circ$C using the following equation:
\begin{equation}
	S(T, f) = S_\mathrm{TETS}(T, f) \times \frac{S_0(23~^\circ\mathrm{C}, f)}{S_\mathrm{TETS}(23~^\circ\mathrm{C}, f) },
\end{equation}
where $S(T,f)$ is the corrected complex sensitivity at temperature $T$ and frequency $f$, and $S_\mathrm{TETS}(T,f)$ and $S_0(T,f)$  are the calibration results obtained with TETS and the existing system, respectively.
Similar correction was applied to Servo 2 and 3 based on their respective comparison measurement.

\begin{figure}[tb]
	\begin{center}
	\begin{minipage}{0.45\columnwidth}
		\centering
		\includegraphics[width=\columnwidth]{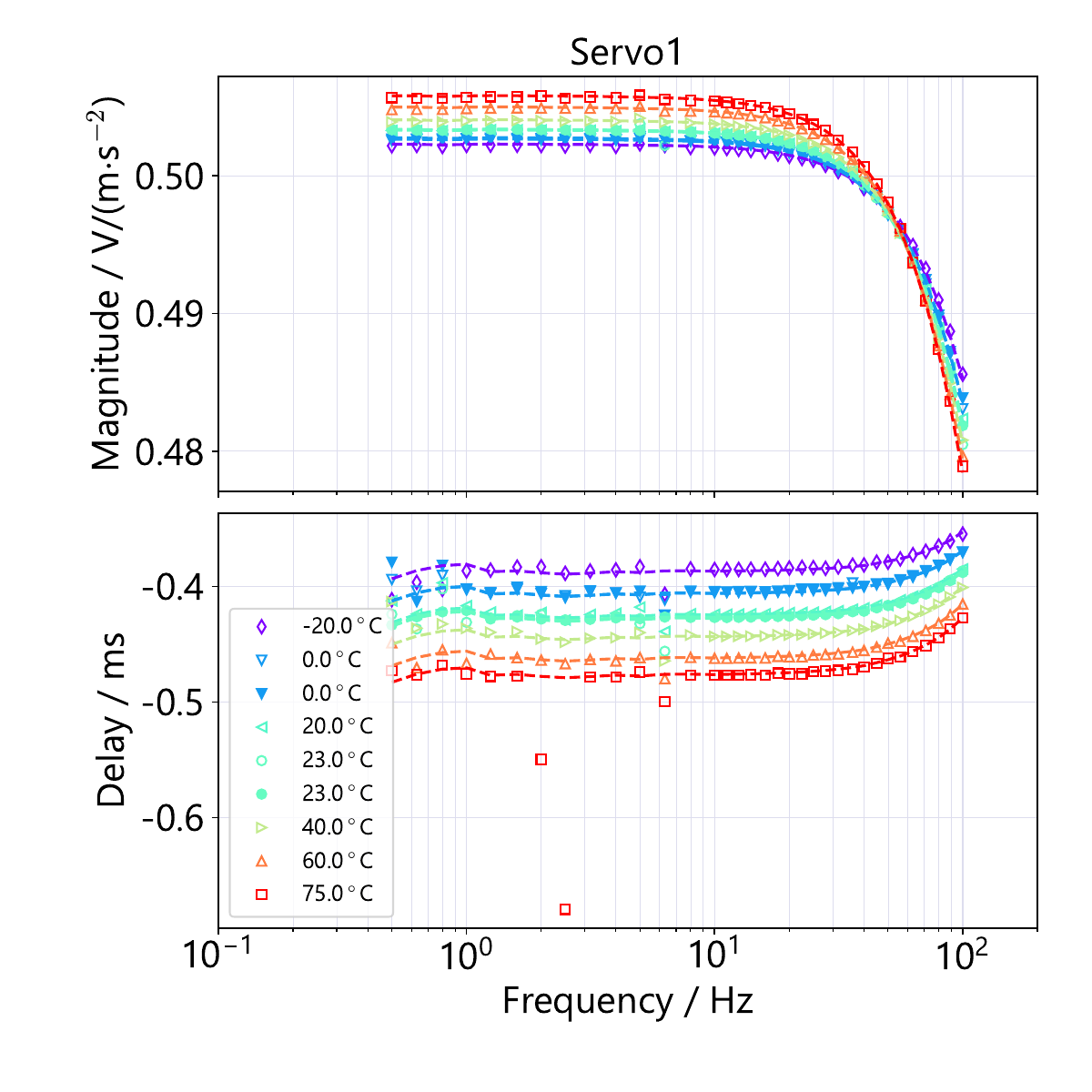}
	\end{minipage}
	\begin{minipage}{0.45\columnwidth}
		\centering
		\includegraphics[width=\columnwidth]{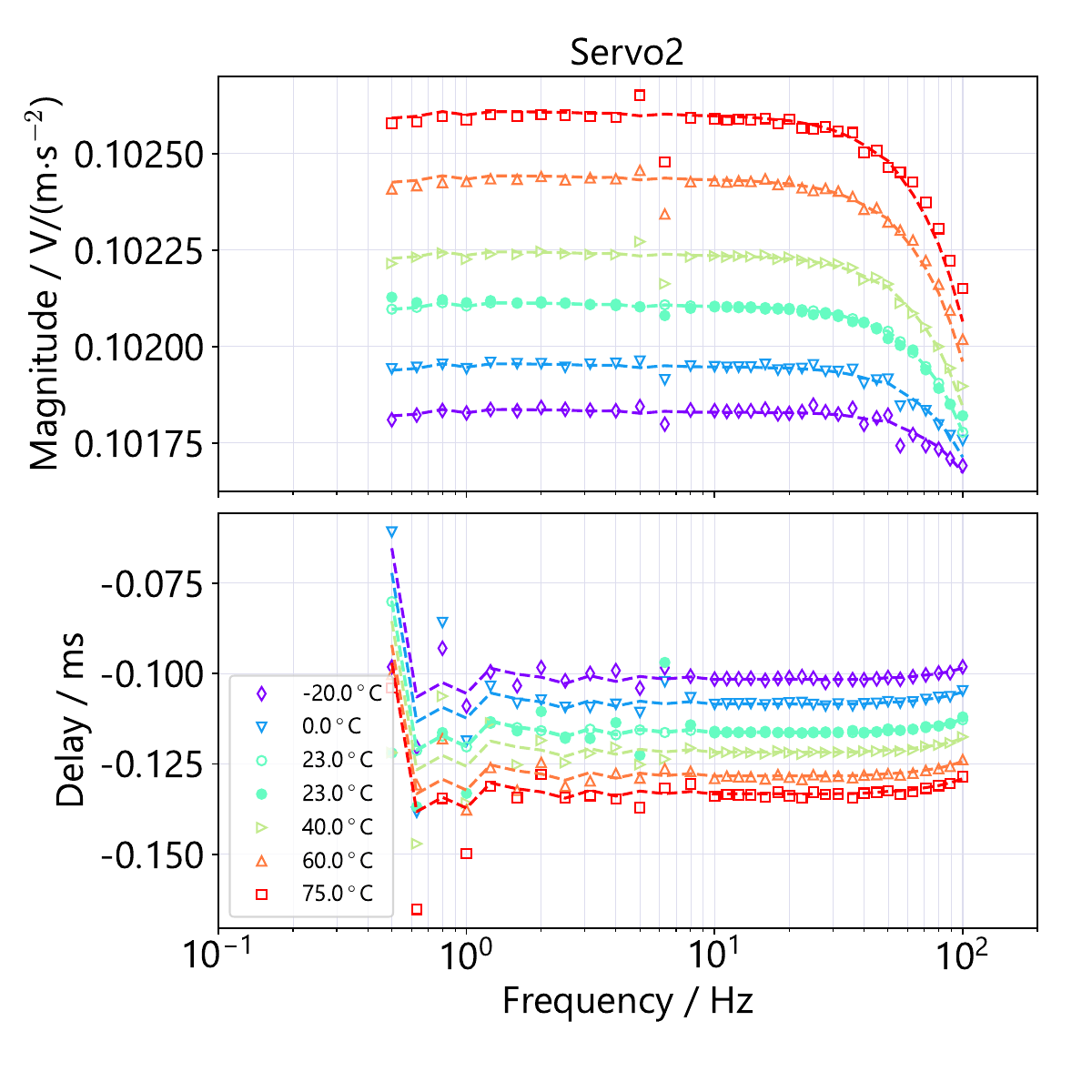}
    	\end{minipage}
	\begin{minipage}{0.45\columnwidth}
		\centering
       	\includegraphics[width=\columnwidth]{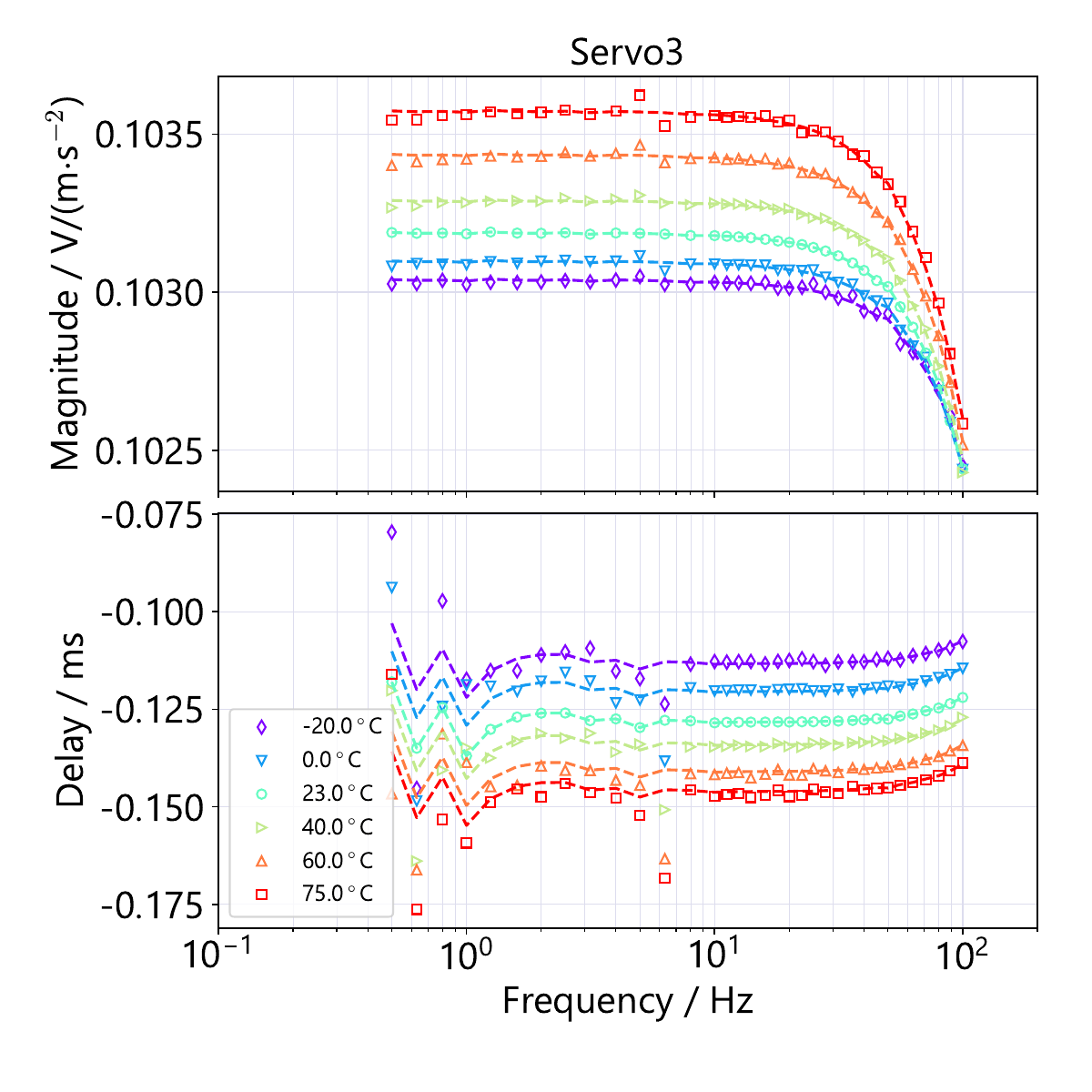}
    	\end{minipage}
	\caption{Primary calibration results at various temperatures for three servo accelerometers. The measured values are plotted with open circles, and the fitting results are plotted with dashed lines.}
	\label{fig:primary_cals}
	\end{center}
\end{figure}
Next, calibration was repeated at different chamber temperatures. 
The temperature was monitored using a thermometer placed near the SUT (Figure~\ref{fig:setup_servo}). 
At each temperature, calibration was conducted one hour after adjusting the airflow temperature to stabilize the SUT. 
During measurements, temperature drift was less than 1.5~$^\circ$C. 
Figure~\ref{fig:primary_cals} shows the temperature dependence of the calibration results for the three servo accelerometers. 
The magnitude of the low-frequency sensitivity shifted by approximately 1~\%, and the delay changed by about 0.04~ms due to variations in the high-frequency cutoff.

To quantitatively evaluate temperature dependence, the complex sensitivity was empirically modeled as:
\begin{equation}
	S(T, f) = S_\mathrm{com}(f) \frac{r(T)}{1 + i f/f_\mathrm{c}(T)},	\label{eq:model}
\end{equation}
where $r(T)$ represents the relative magnitude with respect to 23~$^\circ$C, and $f_\mathrm{c}(T)$ is the cutoff frequency of the first-order low-pass response. 
By definition, $r(23~^\circ\mathrm{C}) = 1$. 
The temperature-independent factor is denoted as $S_\mathrm{com}(f)$. 
Using Eq.~(\ref{eq:model}), the relative sensitivity was fitted as:
\begin{equation}
	\frac{S(T, f)}{S(23~^\circ\mathrm{C}, f)} = r(T) \frac{1 + i f/f_\mathrm{c}(23~^\circ\mathrm{C})}{1 + i f/f_\mathrm{c}(T)}.
\end{equation}
The fitting results, shown in Figure~\ref{fig:primary_cals} with dashed lines, indicate that the model accurately reproduces the measured temperature dependence.
The deviations from the fitted model are likely due to the random background noise of the calibration system or the correction residual of the frame resonance.

\begin{figure}[tb]
	\begin{center}
	\begin{minipage}{0.47\columnwidth}
		\centering
		\includegraphics[width=\columnwidth]{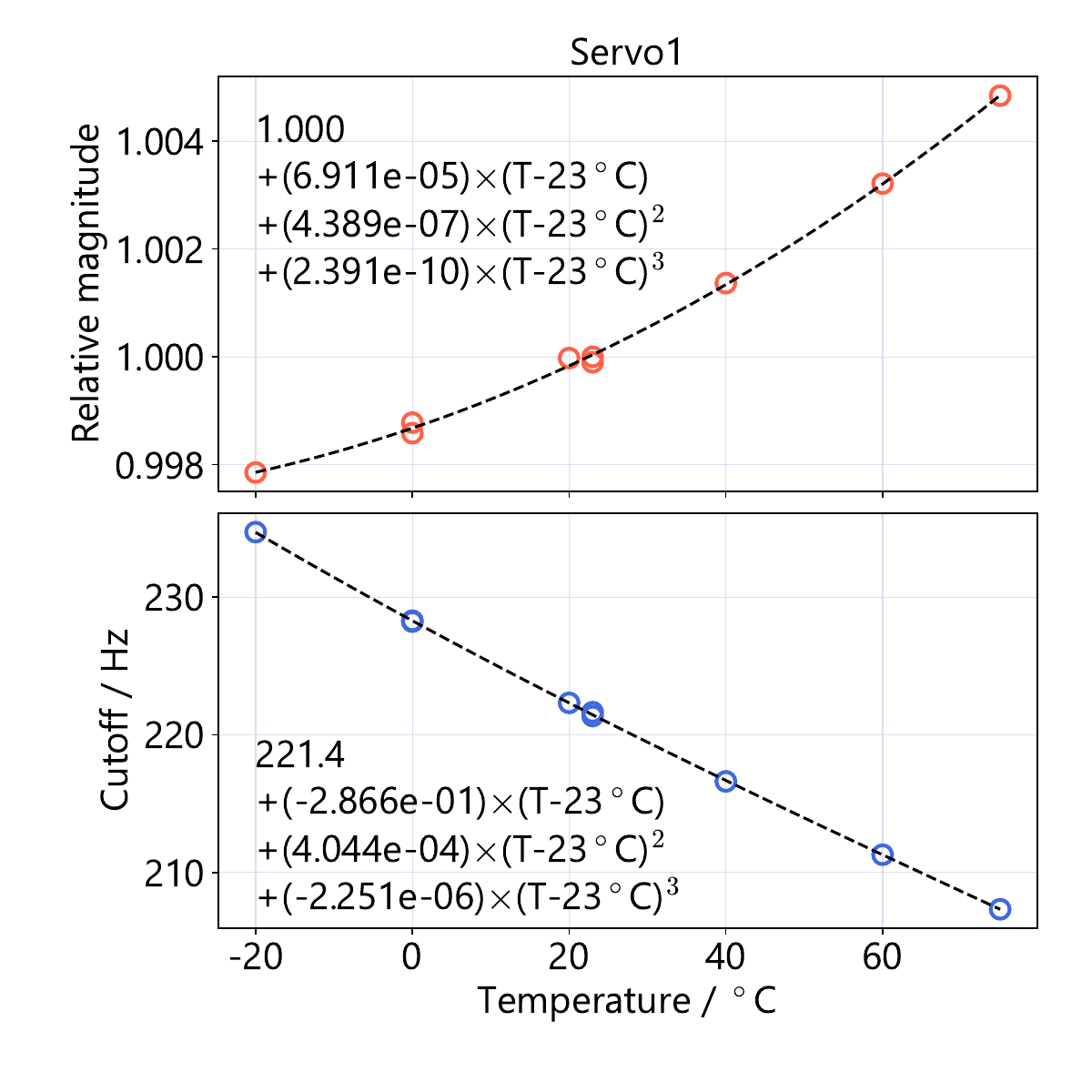}
	\end{minipage}
	\begin{minipage}{0.47\columnwidth}
		\centering
		\includegraphics[width=\columnwidth]{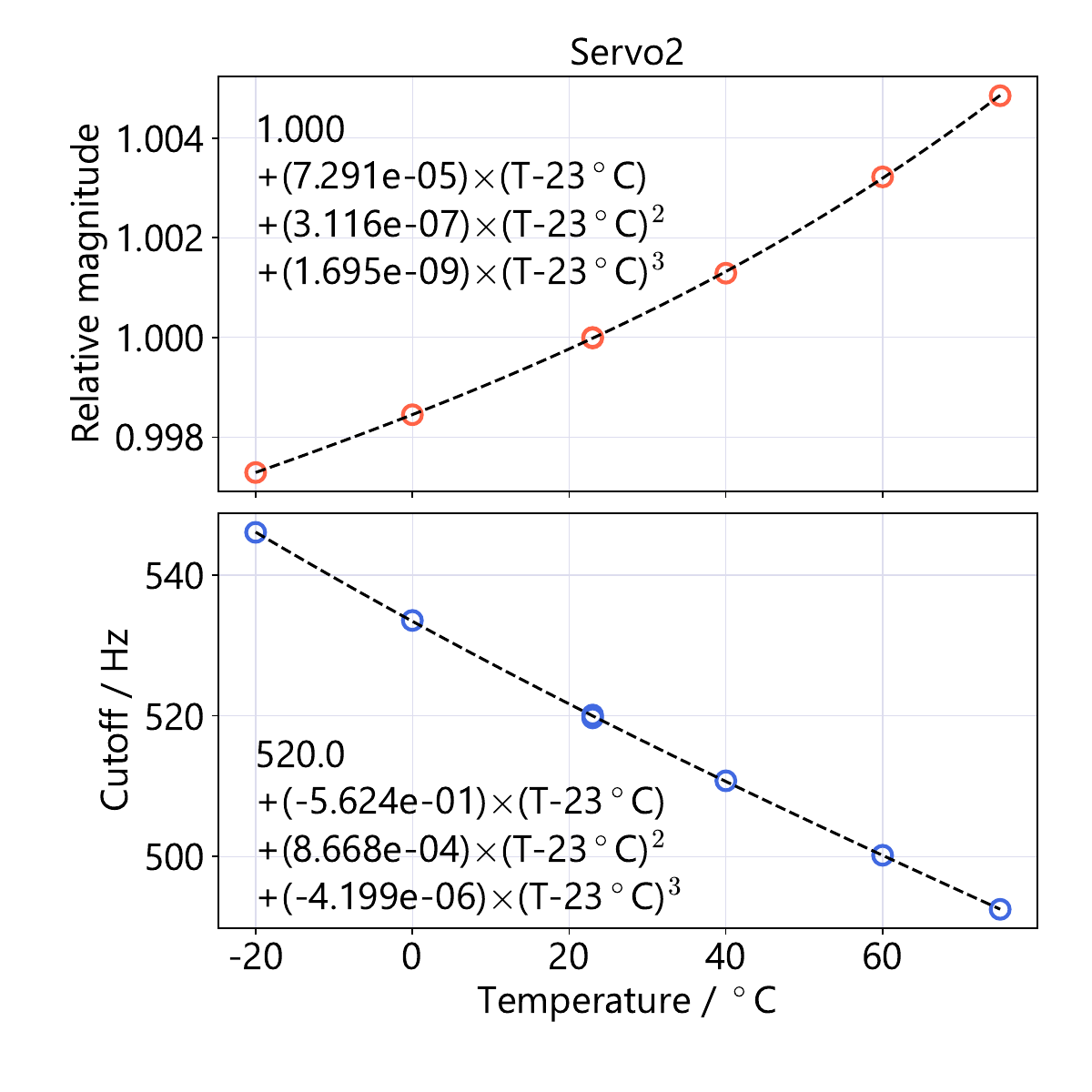}
    	\end{minipage}
	\begin{minipage}{0.47\columnwidth}
		\centering
       	\includegraphics[width=\columnwidth]{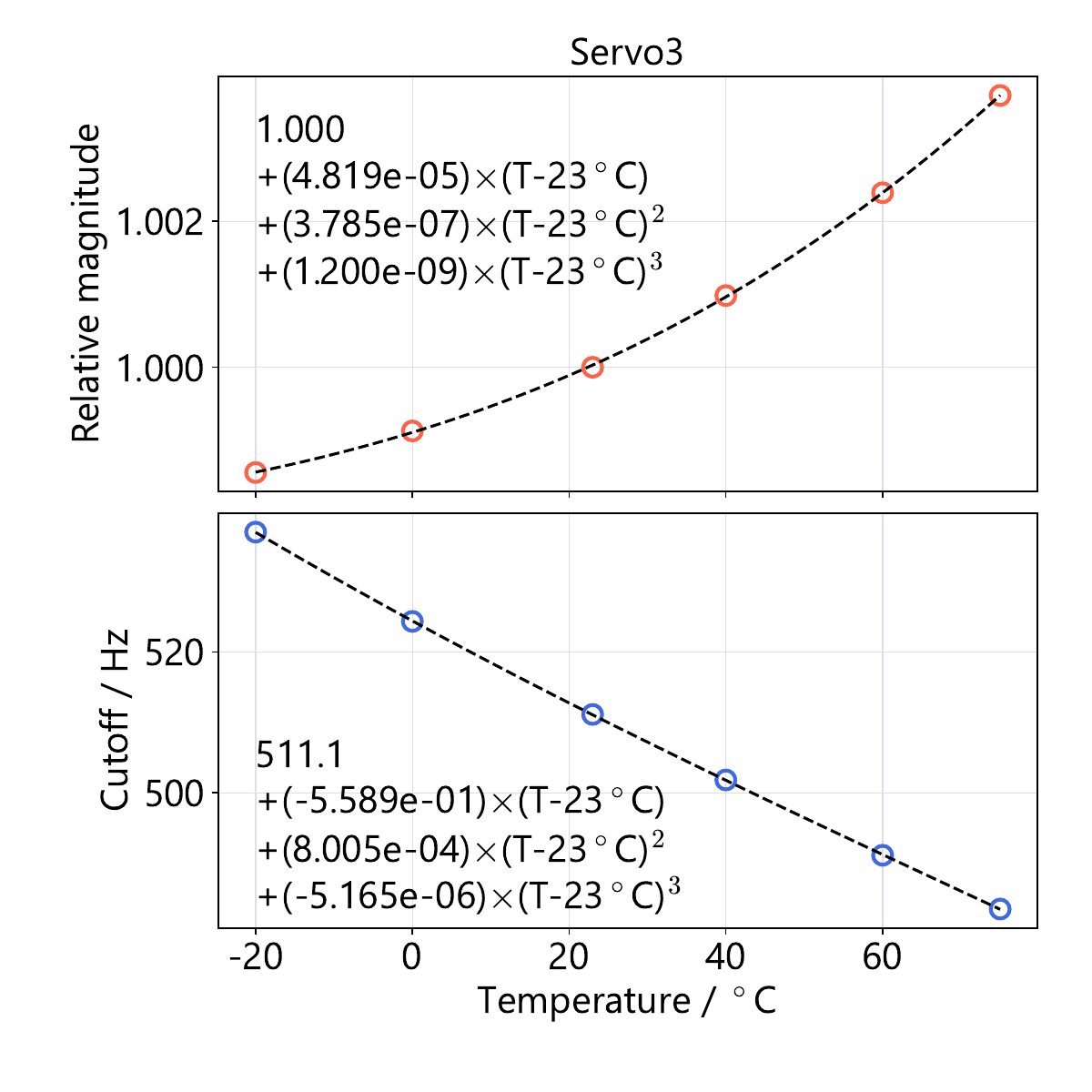}
    	\end{minipage}
	\caption{Temperature dependence of the relative magnitude $r(T)$ and cutoff frequency $f_\mathrm{c}(T)$. The fitting results with a cubic function are shown with dashed lines.}
	\label{fig:servo_params}
	\end{center}
\end{figure}
Furthermore, the parameters $r(T)$ and $f_\mathrm{c}(T)$ were fitted with cubic functions, as shown in Figure~\ref{fig:servo_params}. 
Using these fitted parameters and Eq.~(\ref{eq:model}), sensitivity at any temperature within the range $-20~^\circ\mathrm{C}$ to $+75~^\circ\mathrm{C}$ can be interpolated. 
In subsequent experiments, this interpolated sensitivity was used as the reference for calculating vibration.

The accuracy of the reference sensitivity magnitude is determined by the combination of the primary calibration and the temperature dependence estimation. 
The uncertainty of primary calibration using the existing system is approximately 0.1~\% \cite{Shimoda2023}, mainly due to voltage measurement accuracy, temperature fluctuation, and reproducibility.
The same value is assumed here because the main uncertainty sources are common to TETS. 
Additionally, the deviation from the cubic fit was up to 0.02~\% as observed in Figure~\ref{fig:servo_params}, and temperature measurement uncertainty of about $3~^\circ\mathrm{C}$ contributed 0.03~\% to interpolation uncertainty. 
Combining these factors yields $\sqrt{(0.1~\%)^2+(0.02~\%)^2+(0.03~\%)^2}\approx0.11~\%$, which represents the uncertainty of single-axis reference vibration measurement.

For triaxial vibration, uncertainty increases due to cross-axis coupling, which depends on the waveform. 
The servo accelerometers have a cross-axis sensitivity of approximately 0.2~\%. 
In earthquake simulations, where vibrations along each axis are of similar magnitude, transverse contributions can reach about 0.2~\% per axis.
Therefore, the overall uncertainty of triaxial reference vibration measurement is $\sqrt{(0.11~\%)^2+(0.2~\%)^2}\approx0.23~\%$, which is the sum of the uncertainty of the single-axis measurement and the cross-axis coupling.

\section{Simulation of earthquake observation}	\label{sec:triaxial}
\begin{figure}
	\begin{center}
	\includegraphics[width=0.5\columnwidth]{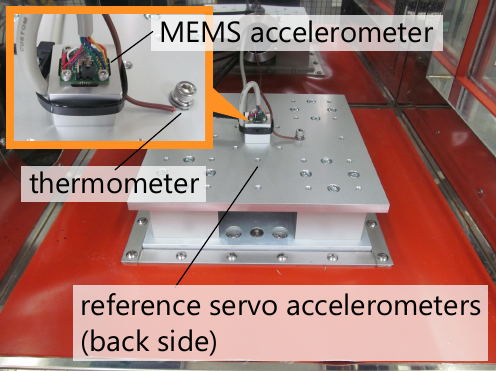}
	\caption{A measurement setup of the servo accelerometer for primary calibration.}
	\label{fig:setup_MEMS}
	\end{center}
\end{figure}
To demonstrate triaxial vibration measurement under varying temperature conditions, an earthquake simulation was performed at $-15~^\circ$C, $+23~^\circ$C, and $+56~^\circ$C. 
A MEMS accelerometer ADXL355 (Analog Devices) was used as the SUT, placed inside the chamber as shown in Figure~\ref{fig:setup_MEMS}. 
Three reference servo accelerometers, calibrated in Section~\ref{sec:primary}, were mounted on the back side of the MEMS accelerometer. 
The vibration waveform recorded during the mainshock of the 2016 Kumamoto earthquakes was applied to the SUT. 
The waveform used was a modified version based on the data provided by KiK-net (Kiban Kyoshin Network) of National Research Institute for Earth Science and Disaster Resilience (NIED) \cite{NIED}.
Temperature dependence of the reference accelerometer sensitivity was corrected when calculating the reference waveform. 
The recorded waveform of the SUT was then compared with the reference. 
As discussed in the previous section, the uncertainty of the reference waveform is approximately 0.23~\%.

\begin{figure*}[tb]
	\begin{center}
	\begin{minipage}{0.32\textwidth}
        \centering
        \includegraphics[width=\columnwidth]{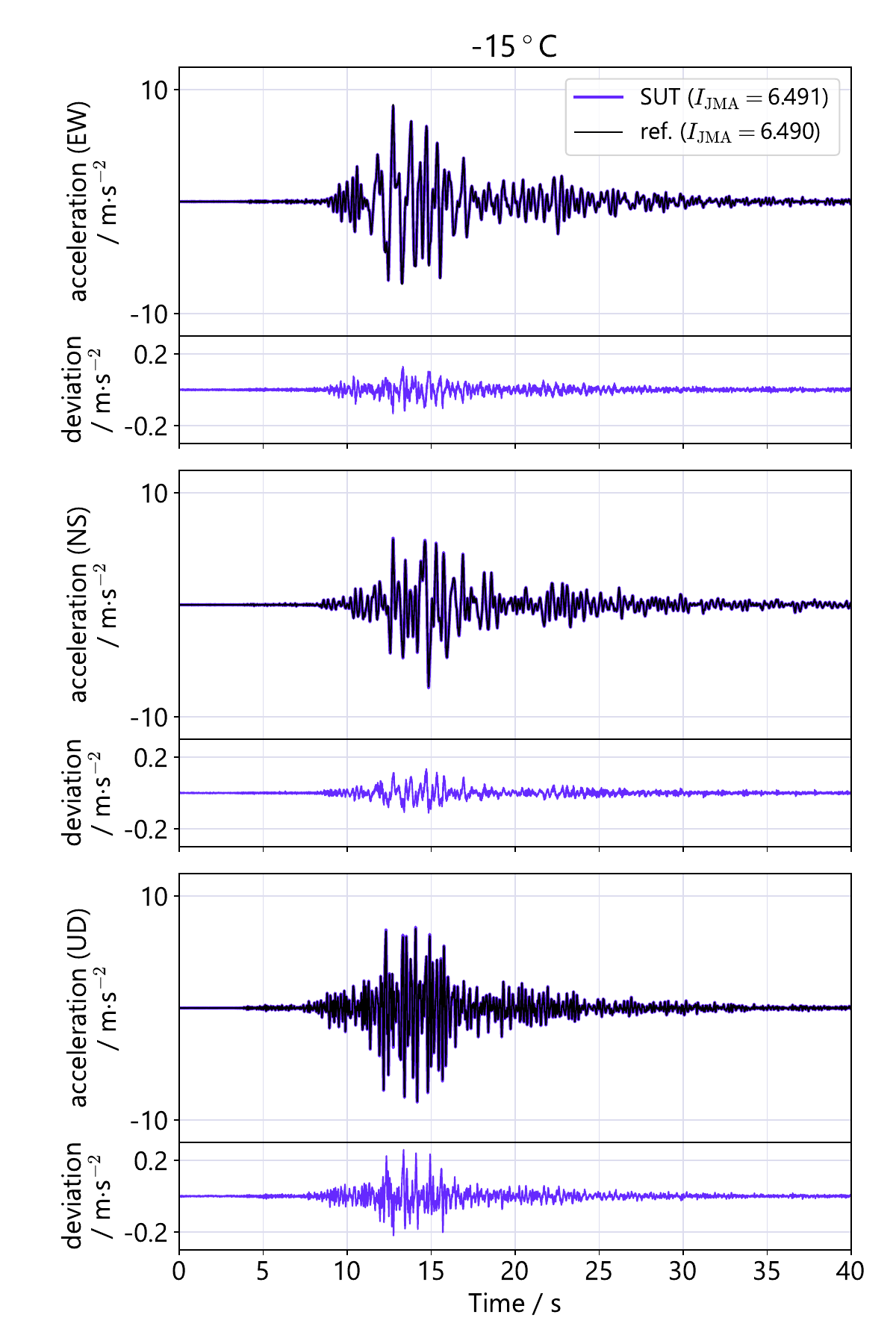}
    \end{minipage}
	\begin{minipage}{0.32\textwidth}
        \centering
        \includegraphics[width=\columnwidth]{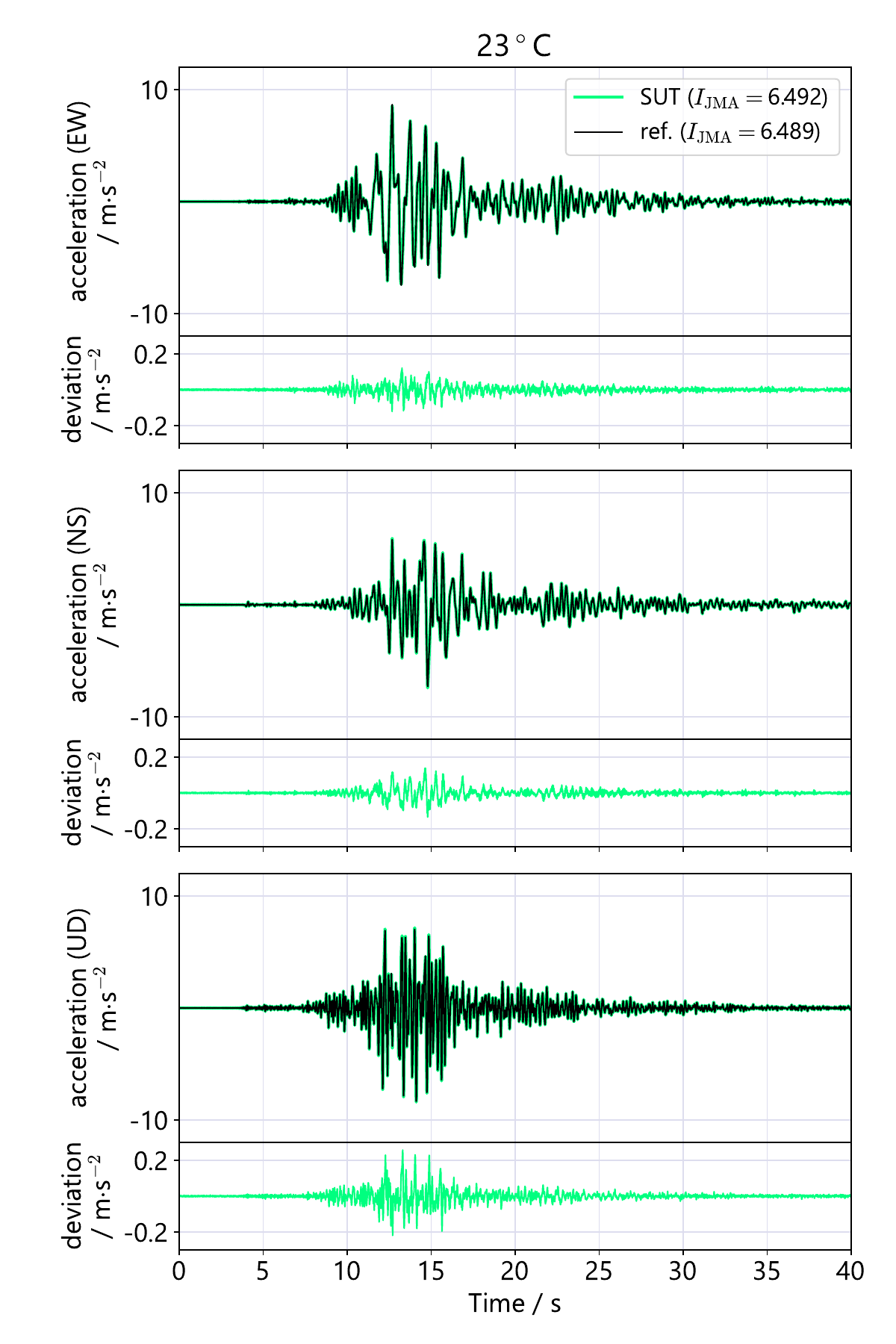}
    \end{minipage}
	\begin{minipage}{0.32\textwidth}
        \centering
        \includegraphics[width=\columnwidth]{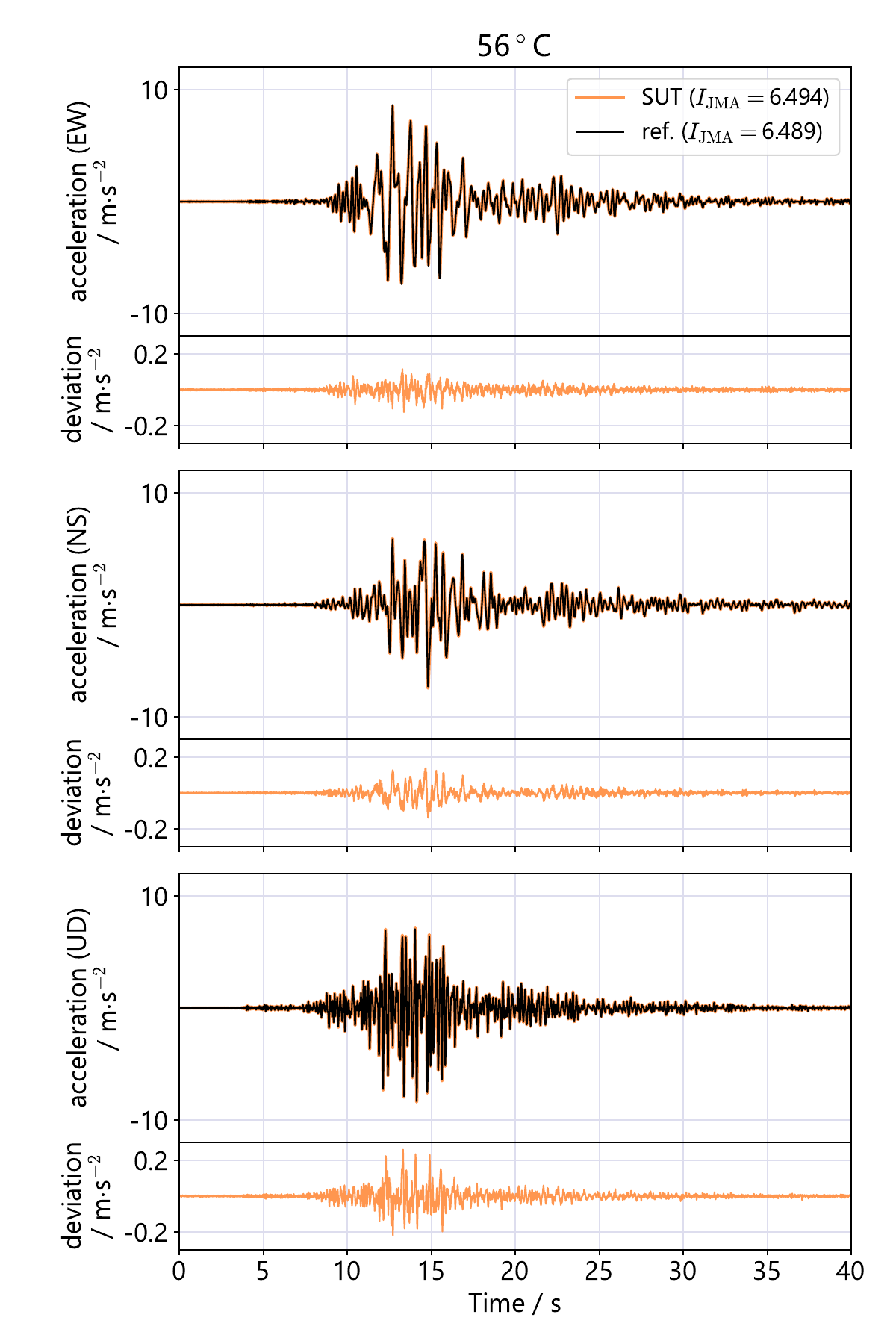}
    \end{minipage}
	\caption{Simulation of earthquake observation at $-15~^\circ$C (left), $23~^\circ$C (center), $56~^\circ$C (right). Each figure shows triaxial vibration waveform (EW, NS, UD) measured by the reference accelerometers (black line) and the SUT (colored line). The difference between the SUT and reference is also plotted for each component. The calculated seismic intensity scale is shown in the legend.}
	\label{fig:triaxial_results}
	\end{center}
\end{figure*}
The simulation results are shown in Figure~\ref{fig:triaxial_results}. 
The waveform deviation---defined as the difference between the reference and SUT waveforms---was up to about 3~\%.
Since this exceeds the reference uncertainty, the deviation is attributed to characteristics of the SUT, such as sensitivity accuracy, cross-axis sensitivity, and waveform distortion due to nonlinearity. 
The results indicate that this deviation is consistent across all tested temperatures.

To evaluate overall measurement performance, the instrumental seismic intensity defined by the Japan Meteorological Agency (JMA), $I_\mathrm{JMA}$, was calculated using triaxial waveforms. 
The calculation procedure follows \cite{Sakai2013}: the recorded waveform is processed using a band-pass-like filter (approximately 0.5~Hz to 10~Hz), and a threshold $a$ is determined such that the duration during which acceleration exceeds $a$ is 0.3~s. 
The seismic intensity is then computed as $I_\mathrm{JMA}=2\log_{10} a+0.94$.
Although $I_\mathrm{JMA}$ is typically rounded to one decimal place, values up to three decimal places were used here to examine small variations.

The calculated seismic intensities for the SUT and the reference are shown in Figure~\ref{fig:triaxial_results}. 
The reference uncertainty of approximately 0.23~\% corresponds to an uncertainty of about 0.002 in seismic intensity. 
The calculated $I_\mathrm{JMA}$ from the SUT exhibited larger deviations from the reference at higher temperatures, likely due to temperature-dependent sensitivity changes. 
Nevertheless, the SUT demonstrated sufficient accuracy for determining seismic intensity when rounded to one decimal place.

\section{Conclusion} \label{sec:conclusion}
A triaxial environmental test system was developed by integrating a triaxial vibration exciter with a thermostatic chamber. 
This study focused on evaluating the effect of temperature on vibration measurement accuracy. 
Reference vibrations were measured using servo accelerometers that were primarily calibrated at various temperatures, enabling accurate measurement of long-stroke, low-frequency vibrations such as earthquakes with an overall uncertainty of approximately 0.23~\%. 
The primary calibration was also performed within the TETS, allowing the entire measurement process to be completed using a single system.
As a demonstration, a MEMS accelerometer was tested between $-15~^\circ$C and $+56~^\circ$C, and the triaxial vibration measurement process was validated. 
The results confirmed that the SUT maintained sufficient accuracy for seismic observation.

Similar tests under various environmental conditions can improve the reliability of vibration measurements. 
With the growing demand for field measurements in disaster prevention, such as earthquake monitoring and infrastructure health assessment, ensuring data reliability is critical because measurement results ultimately impact human safety.
Therefore, systematic evaluation of measurement accuracy is essential and will contribute to the advancement and dissemination of vibration sensing technologies.

\section*{Acknowledgment}
This work was supported by a project commissioned by the New Energy and Industrial Technology Development Organization (NEDO), Japan. 
The waveform used in the earthquake simulation was based on the data obtained by KiK-net of NIED \cite{NIED}.

\section*{Conflict of Interest Statement}
The authors have no conflicts to disclose.

\section*{Author Contributions}
\textbf{T. Shimoda:} Conceptualization, Methodology, Investigation, Formal analysis, Resources, Software, Visualization, Funding acquisition, Writing – original draft
\textbf{W. Kokuyama:} Writing – review and editing
\textbf{H. Nozato:} Conceptualization, Methodology, Supervision, Project administration, Funding acquisition, Writing – review and editing

\section*{Data Availability Statement}
The data that support the findings of this study are available from the corresponding author upon reasonable request.

\section*{References}
\providecommand{\noopsort}[1]{}\providecommand{\singleletter}[1]{#1}%


\begin{thebibliography}{1}

\bibitem{Martin2017}
Martin Iwanczik, Philipp Begoff, Michael Mende, and Holger Nicklich.
\newblock Traceable calibration of vibration sensors in a wide frequency and
  temperature range.
\newblock {\em IMEKO 4th TC22 International Conference}, 2017.

\bibitem{Zhu2018}
G.~Zhu, X.~W. Yang, and X.~Liu.
\newblock Calibration of accelerometer temperature sensitivity by laser
  interferometer.
\newblock {\em Journal of Physics: Conference Series}, 1065(22):222003, 2018.

\bibitem{ISO16063-34}
International Organization for Standardization.
\newblock {\em ISO 16063-34:2019, Methods for the calibration of vibration and
  shock transducers - Part 34: Testing of sensitivity at fixed temperatures},
  2019.

\bibitem{ISO16063-11}
International Organization for Standardization.
\newblock {\em ISO 16063-11:1999, Methods for the calibration of vibration and
  shock transducers - Part 11: Primary vibration calibration by laser
  interferometry}, 1999.

\bibitem{Shimoda2023}
T.~Shimoda, W.~Kokuyama, and H.~Nozato.
\newblock Primary microvibration standards down to $10^{-3}$ m s$^{-2}$ at low
  frequency.
\newblock {\em Measurement Science and Technology}, 34(9):095003, 2023.

\bibitem{NIED}
{NIED K-NET, KiK-net, National Research Institute for Earth Science and
  Disaster Resilience}.
\newblock doi:10.17598/NIED.0004.

\bibitem{Sakai2013}
A.~Sakai.
\newblock A method of expressing seismic intensity for a wider period range.
\newblock {\em Journal of JSCE}, 1(1):262, 2013.

\end{thebibliography}
\end{document}